\documentclass[aip,reprint]{revtex4-1}

\usepackage{graphicx}
\usepackage{amsmath}
\usepackage{textcomp} 
\usepackage{xcolor}
\usepackage{ulem}
\begin{document}
\newcommand{\g}{$g_\mathrm{eff}^{\uparrow \downarrow}$}

\title{Optimization of the yttrium iron garnet/platinum interface for spin pumping-based applications}

\author{M.~B.~Jungfleisch}
\email{jungfleisch@physik.uni-kl.de}
\noaffiliation
\affiliation{Fachbereich Physik and Landesforschungszentrum OPTIMAS, Technische Universit\"at Kaiserslautern, 67663
Kaiserslautern, Germany}

\author{V.~Lauer}
\noaffiliation
\affiliation{Fachbereich Physik and Landesforschungszentrum OPTIMAS, Technische Universit\"at Kaiserslautern, 67663
Kaiserslautern, Germany}

\author{R.~Neb}
\noaffiliation
\affiliation{Fachbereich Physik and Landesforschungszentrum OPTIMAS, Technische Universit\"at Kaiserslautern, 67663
Kaiserslautern, Germany}

\author{A.~V.~Chumak}
\noaffiliation
\affiliation{Fachbereich Physik and Landesforschungszentrum OPTIMAS, Technische Universit\"at Kaiserslautern, 67663
Kaiserslautern, Germany}

\author{B.~Hillebrands}
\noaffiliation
\affiliation{Fachbereich Physik and Landesforschungszentrum OPTIMAS, Technische Universit\"at
Kaiserslautern, 67663 Kaiserslautern, Germany}

\date{\today}

\begin{abstract}

The dependence of the spin pumping efficiency and the spin mixing conductance on the surface processing of yttrium iron garnet (YIG) before the platinum (Pt) deposition has been investigated quantitatively. The ferromagnetic resonance driven spin pumping injects a spin polarized current into the Pt layer, which is transformed into an electromotive force by the inverse spin Hall effect. Our experiments show that the spin pumping effect indeed strongly depends on the YIG/Pt interface condition. We measure an enhancement of the inverse spin Hall voltage and the spin mixing conductance of more than two orders of magnitude with improved sample preparation.


\end{abstract}

\maketitle

In the last decades, there was rapidly increasing interest in the field of spintronics. The
promising aim is to exploit the intrinsic spin of electrons to build efficient magnetic storage
devices and computing units. \cite{Zutic} An emerging sub-field of spintronics is magnon spintronics, where
magnons, the quanta of spin waves (collective excitations of coupled spins in a magnetically
ordered solid), are used to carry and process information. \cite{Kajiwara} Magnons pose a number of
advantages to conventional spintronics. One of  them is the realization of insulator-based devices with
decreased energy consumption: Since spin-wave based spin currents in insulators are not accompanied by
charge currents, parasitic heating due to the movement of electrons can be excluded. \cite{Serga}

In order to combine magnon spintronics and charge-based electronics, it is necessary to create
effective converters, which transform spin-wave spin currents into conventional charge currents. The
combination of the spin pumping \cite{Tserkovnyak, Costache} and the inverse spin Hall effect (ISHE)  \cite{Hirsch, Saitoh-2006} turned out to be an
excellent candidate for this purpose. Spin pumping refers to the generation of spin polarized
electron currents in metals by the magnetization precession in an adjacent ferromagnetic layer,
whereas the ISHE transforms this spin current into a conventional
charge current.

In the last years, hetero-structures consistent of a magnetic insulator yttrium iron garnet
(YIG) film and an adjacent platinum (Pt) layer \cite{Kajiwara} attracted considerable attention. Since YIG is an
insulator with a band gap of 2.85~eV \cite{Jia} no direct transition of spin polarized electron
currents from the YIG into the Pt layer is possible. Thus, spin pumping is the only method
applicable in these structures to inject spin currents into the Pt layer. It has been shown, that
standing \cite{Jungfleisch} as well as propagating \cite{Chumak} magnons in a wide range of
wavelengths from centimeters to hundred nanometers \cite{Sandweg2,Kurebayashi} can be efficiently
converted into charge currents using a combination of spin pumping and ISHE. Furthermore, the Pt thickness dependence on  the ISHE voltage from spin pumping \cite{Castel} and non-linear spin pumping \cite{Ando_para} have been investigated.

Since spin pumping is an interface effect, it is of crucial importance to investigate how to control and manipulate the YIG/Pt interface condition in order to obtain an optimal magnon to spin current conversion efficiency. Recently, the influence of Ar$^{+}$ ion beam etching on the spin pumping efficiency in YIG/Au
structures was investigated. \cite{Burrowes} It was shown, that the spin mixing conductance determined by the Gilbert damping constant can be increased by a factor of 5 using Ar$^{+}$ etching. Nevertheless, there are no systematic, quantitative
studies of the influence of the YIG/Pt interface treatment on the ISHE voltage from spin pumping up
to now.

\begin{figure}[b]
\includegraphics[width=0.97\columnwidth]{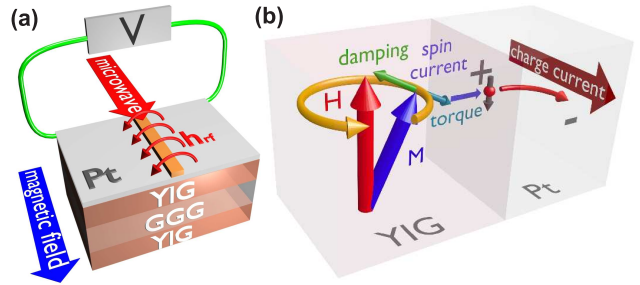}
\caption{\label{fig:setup} (Color online) (a) Sketch of the sample and the setup geometry. (b) Illustration of spin pumping and inverse spin Hall effect. Details see text.}
\end{figure}

\begin{figure*}[t!]
\includegraphics[width=1.6\columnwidth]{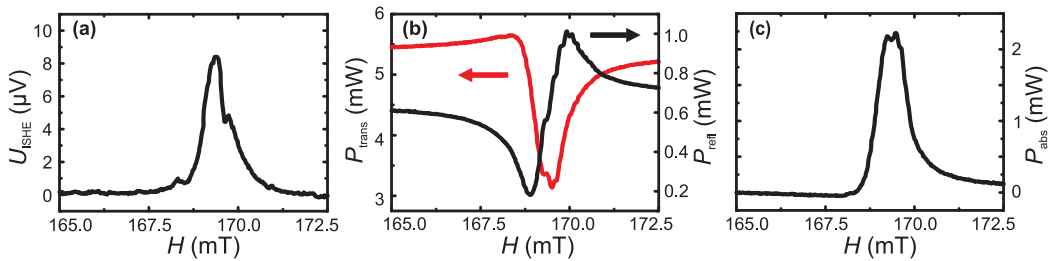}
\caption{\label{fig2} (Color online) (a) ISHE voltage as a function of the applied magnetic field $H$. Magnetic field dependence of (b) transmitted and reflected microwave power $P_\mathrm{trans}$, $P_\mathrm{refl}$ and (c) absorbed microwave power $P_\mathrm{abs}$. Applied microwave
power $P_\mathrm{applied} = 10$ mW, YIG thickness: 2.1 \textmu m.}
\end{figure*}

In this Letter, we present our results on the influence of processing of the YIG film surface before the Pt deposition on the spin pumping efficiency. 
We measure the ferromagnetic resonance (FMR) spectra using conventional microwave techniques, as well as the inverse spin Hall voltage, which allows us to calculate the spin pumping efficiency, defined as the ratio of the detected ISHE charge current to the absorbed microwave power. Our experimental results clearly show a significant difference (up to factor of 150) between the different surface treatments.

A sketch of the experimental setup is shown in Fig.~\ref{fig:setup}(a). The YIG samples of 2.1 \textmu m and 4.1 \textmu m thickness and a size of $3\times 4$ mm$^2$ were grown by liquid phase epitaxy on both sides of 500 \textmu m thick gadolinium gallium garnet (GGG) substrates.
Since the GGG substrate between the two YIG layers is rather thick  (500 \textmu m), the second YIG layer has no influence on our studies.

After a conventional pre-cleaning step, all samples were cleaned by acetone and isopropanol in an ultrasonic bath. In order to provide the same cleanliness of all samples, the purity of each sample was monitored after this first step.  Afterwards, the following surface treatments have been applied:

\begin{itemize}

    \item Cleaning in ``piranha'' etch, a mixture of H$_{2}$SO$_{4}$ and H$_{2}$O$_{2}$. 
    This exothermic reaction at around 120$^\circ$C is strongly oxidizing and, thus, removes most organic matter. Among the used surface treatments, ``piranha'' cleaning is the only one, which was performed outside the molecular beam epitaxy (MBE) chamber. Thus, we cannot exclude, that the samples are contaminated by microscopic dirt or a water film due to air exposure before the Pt deposition.

    \item Heating at 200$^\circ$C for 30 min in order to remove water from the sample surfaces, performed \textit{in-situ}.
    \item Heating at 500$^\circ$C for 5 hours performed \textit{in-situ} in order to remove water  from the sample surfaces. It might be that for this method lattice misfits in the YIG crystal are annealed as well. \cite{Choi, Ramer}
    \item \textit{In-situ} Ar$^{+}$ plasma cleaning at energies of 50-100 eV for 10 min. SRIM simulations show that the used energies are below the threshold for sputtering and, thus, this cleaning method acts mechanically. 
    \item \textit{In-situ} O$^{+}$/Ar$^{+}$ plasma cleaning 
    at energies around 50-100 eV for 10 min. In addition to the mechanical cleaning effect, O$^{+}$ oxidizes organic matter.
 \end{itemize}

\hspace{-10pt}The used surface processings are summarized in Tab.~\ref{tab}. They fulfill different requirements: First of all, the samples are cleaned removing microscopic particles, (organic) matter and water. In the case of the samples, which are heated, the crystalline YIG structure might be annealed as well. Another aspect of the used treatments might be the modification of the spin pinning condition (not part of the present studies; requires further deeper investigations). After cleaning, the 10 nm thick Pt layer was grown by MBE at a pressure of $5\times10^{-8}$ mbar and a growth rate of 0.01 nm/s. 
 It is important to note that the Pt film was deposited on each of the samples of one set simultaneously, ensuring identical growth conditions.

The measurement of the spin pumping efficiency was performed in the following way. The samples were magnetized in the film plane by an external magnet field $\textbf{H}$  (see Fig.~\ref{fig:setup}(a)). The magnetization precession was excited at a constant frequency of $f = 6.8$ GHz by applying microwave signals of power $P_\mathrm{applied}$ to a  600 \textmu m wide Cu microstrip antenna. 
The Pt layer and the microstrip antenna were electrically isolated by a silicon oxide layer of 100 \textmu m thickness in order to avoid overcoupling of the YIG film with the antenna. While sweeping the external magnetic field  the inverse spin Hall voltage $U_\mathrm{ISHE}$ (Fig.~\ref{fig2}(a))  as well as the microwave reflection and transmission (Fig.~\ref{fig2}(b)) were recorded. The voltage $U_\mathrm{ISHE}$ was measured across the edges of the Pt layer perpendicular to the external magnetic field using a lock-in technique. For this purpose the microwave amplitude was modulated with a frequency of 500 Hz. Changing the external magnetic field to the opposite direction results in an inverted  voltage proving the ISHE nature of the observed signal. \cite{Hirsch,Saitoh-2006,Kajiwara}
The complicated absorption and reflection spectra depicted in Fig.~\ref{fig2}(b) are due to the interference of the electromagnetic signal in the microstrip line reflected from the YIG sample and the edges of the line. \cite{Barry}
This behavior does not influence our studies since we further use only the maximum of  $U_\mathrm{ISHE}$ to calculate the spin pumping efficiency (see Fig.~\ref{fig2}(c)). Knowing the applied microwave power $P_\mathrm{applied}$ and measuring microwave reflection $P_\mathrm{refl}$ and transmission $P_\mathrm{trans}$ (see Fig.~\ref{fig2}(b)) enables us to calculate the absorbed power as $P_\mathrm{abs}=P_\mathrm{applied}-(P_\mathrm{trans}+P_\mathrm{refl}$). The results are depicted in Fig.~\ref{fig2}(c). \cite{modes} At the FMR field $H_\mathrm{FMR} \approx 170$ mT, energy is transferred most effectively into the magnetic system and, thus, the microwave absorption is maximal (see Fig.~\ref{fig2}(c)). In resonance condition, the angle of precession is maximal and spin currents are most efficiently pumped from the YIG into the Pt layer (Fig.~\ref{fig:setup}(b)) and transformed into an electric current by the ISHE. \cite{Saitoh-2006,Kajiwara} Subsequently, we measure the maximal ISHE voltage $U_\mathrm{ISHE}^\mathrm{FMR}$ at $H_\mathrm{FMR}$.

\begin{figure}[t]
\includegraphics[width=0.72\columnwidth]{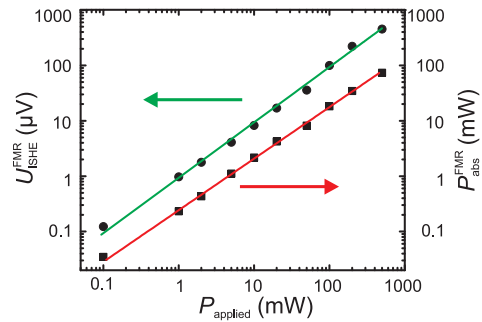}
\caption{\label{fig4} (Color online) Maximal inverse spin Hall effect induced voltage $U_\mathrm{ISHE}^\mathrm{FMR}$ and absorbed microwave power $P_\mathrm{abs}$ at FMR as a function of the applied microwave power $P_\mathrm{applied}$. YIG thickness: 2.1 \textmu m, cleaning method 7, ``piranha'' etch and heating at 500$^{\circ}$C.}
\end{figure}

The dependence of  $U_\mathrm{ISHE}^\mathrm{FMR}$ as a function of the applied microwave power $P_\mathrm{applied}$ is shown in Fig.~\ref{fig4}, left scale (illustrated is method 7, Tab.~\ref{tab}, discussed below with a rather high enhancement factor of 104). The interface optimization provides the possibility to observe  $U_\mathrm{ISHE}^\mathrm{FMR}$ over wide range of applied powers  $P_\mathrm{applied}$. We find a linear relation between $U_\mathrm{ISHE}^\mathrm{FMR}$  and $P_\mathrm{applied}$ over the whole power region of nearly four  orders of magnitude. The ISHE voltage $U_\mathrm{ISHE}^\mathrm{FMR}$ increases from 100 nV (for $P_\mathrm{applied}\approx 100~$\textmu W) to approximately 500 \textmu V (for $P_\mathrm{applied}\approx 500$ mW). On the right scale in Fig.~\ref{fig4} the absorbed microwave power $P_\mathrm{applied}$ is shown as function of the applied microwave power $P_\mathrm{abs}$. $P_\mathrm{abs}$ depends also linearly on $P_\mathrm{applied}$. 

In order to investigate the different cleaning methods described above we measure the spectra for three different microwave powers $P_\mathrm{applied}$ of 1 mW, 10 mW and 100 mW. We investigate  three sets of samples (2 sets at 2.1 \textmu m, 1 set at 4.1 \textmu m YIG thickness).

In order to compare the samples of one set we introduce the spin pumping efficiency as

\begin{equation}
\label{efficiency}
\eta(P_\mathrm{applied},n)=\frac{U_\mathrm{ISHE}^{\mathrm{FMR}}}{R \cdot P_\mathrm{abs}^{\mathrm{FMR}}},
\end{equation}

\hspace{-13pt} where $n$ is the index number of the sample (i. e. the index number of the cleaning method, see Tab.~\ref{tab}), $U_\mathrm{ISHE}^{\mathrm{FMR}}$ is the maximal ISHE voltage in resonance $H_\mathrm{FMR}$, $R$ is the electric resistance of the Pt layer, $P_\mathrm{applied}$ is the applied microwave power and $P_\mathrm{abs}^{\mathrm{FMR}}$ is the absorbed microwave power at  $H_\mathrm{FMR}$.

Further, we introduce the power and thickness independent parameter $\epsilon$ by normalizing the efficiency of the $n$-th sample to the first sample of each set $n=1$, which underwent only a simple cleaning process, as

\begin{equation}
\label{epsilon}
\epsilon(n)=\frac{\eta(n,P_\mathrm{applied})}{\eta(n=1,P_\mathrm{applied})}.
\end{equation}

The $\epsilon$-parameter is a measure for the enhancement of the spin pumping efficiency due to the surface treatment. For each of them we calculate the $\epsilon$-values: three microwave powers for each set. The general tendency is the same for all series of measurements.  
According to the literature \cite{Nakayama-2012}, the spin pumping efficiency should not dependent on the YIG thickness for the used samples due to their large thicknesses. We observe an ISHE voltage for the 4.1 \textmu m set to be around 20 \% of  that of the 2.1 \textmu m sets, which we associate with a better quality of the 2.1 \textmu m YIG film. Nevertheless, the general tendency of the enhancement $\epsilon(n)$ due to the used surface treatments is rather independent on the film thickness: the absolute value of the $\epsilon$-parameter for the 4.1 \textmu m set is 80 \% of that of the  2.1 \textmu m sets. 
In order to obtain an easily comparable measure for the spin pumping efficiency, we further introduce the mean value $\bar{\epsilon}$ of these  $\epsilon$-values. 
The standard deviation is given by $\sigma(\bar{\epsilon})$ and is a measure for the reliability of the specific surface treatments. The results are summarized in Tab.~\ref{tab}.

Our investigations of the YIG/Pt  interface optimization on spin pumping show the following trend.
The conventional cleaning by acetone and isopropanol in an ultrasonic bath (sample $1$) achieves the worst results ($\bar{\epsilon} = 1$, $U_\mathrm{ISHE} = 270$ nV for $P_\mathrm{applied} = 10$ mW). Surface cleaning by ``piranha'' improves the efficiency by a factor of 14, but this method is not reliable (standard deviation of $\sigma(\bar{\epsilon}) = 131 \%$ is very large). 
Since this surface treatment takes place outside the MBE chamber and since the sample is in air contact after cleaning, the sample might be contaminated again by water and possibly by microscopic dirt before the Pt deposition. Heating the samples after ``piranha'' cleaning in the MBE chamber at  $200^\circ$C for 30 min (sample $3$) removes mainly the water film from the surface and results in a 64 times higher efficiency. However, this cleaning method also does not guarantee a high ISHE voltage, which is reflected in the high standard deviation. Using an Ar$^{+}$ plasma (method $4$) is more efficient (see Fig.~\ref{fig2}) and particularly more reliable. Method $4$ acts mechanically and removes water as well as other dirt from the sample surface. The best results ($\bar{\epsilon} = 152$) are obtained for the O$^{+}$/Ar$^{+}$ plasma (sample $5$). The additional advantage of this surface treatment is the removal of organic matter. In order to check the reliability of the O$^{+}$/Ar$^{+}$ plasma cleaning (method $5$), we substituted the ``piranha'' cleaning by heating the sample: even without ``piranha'' etching, but with heating at  $200^\circ$C and O$^{+}$/Ar$^{+}$ plasma (method $6$), we achieve a considerable spin pumping efficiency of $\bar{\epsilon} = 86$. This is mainly attributed to the O$^{+}$ plasma. It is remarkable that purely heating the sample at $500^\circ$C for 5 hours (method $7$) results in a comparable high efficiency of $\bar{\epsilon} = 104$. The additional reason might be, that the temperature is sufficiently high to anneal crystal defects of the YIG samples. \cite{Ramer, Choi}

\begin{table}
\centering
\begin{tabular}[b]{c l r r  c}
\hline \hline
\multicolumn{1}{c}{sample $n$} & \multicolumn{1}{c}{process} & \multicolumn{1}{c}{$\bar{\epsilon}$} &  \multicolumn{1}{c}{$\sigma(\bar{\epsilon})$ [\%]} & \g  ($\times 10^{19}$ m$^\mathrm{-2}$) \\

\hline 
1 & simple cleaning &  1 &  --- & 0.02\\
2 & piranha & 14 & 131 & 0.32\\
3 & piranha + 200$^\circ$C & 64 & 93 & 1.45\\
4 & piranha + Ar$^{+}$  & 79 & 54 & 1.79\\
5 & piranha + $\mathrm{O}^{+}$/Ar$^{+}$  & 152 & 61 & 3.43\\
6 & 200$^\circ$C + $\mathrm{O}^{+}$/Ar$^{+}$  & 86 & 52 & 1.94\\
7 & piranha + 500$^\circ$C & 104 & 40 & 2.35\\
\hline \hline
\end{tabular}
\caption{\label{tab}$\bar{\epsilon}$ is the calculated mean value of the $\epsilon$-parameter, which is the thickness and power independent spin pumping efficiency (Eq.~\ref{epsilon}), 
$\sigma$($\bar{\epsilon}$) is the standard deviation, spin mixing conductance \g.}
\end{table}

In order to determine the spin mixing conductance \g\ of our samples, we performed FMR measurements on one of the YIG samples with a pronounced mode structure (method 7, with and without Pt layer on the top) using a vector network analyzer. \cite{FMR} The FMR linewidth $\Delta H$ is related to the Gilbert damping parameter as\cite{Burrowes,Nakayama-2012,Mosendz,Heinrich} $\alpha = \gamma \Delta H/{2 \omega}$,
where $\gamma$ is the gyromagnetic ratio and $\omega=2\pi f$ is the microwave angular frequency.
For the bare YIG sample we measure $\Delta H_\mathrm{0}= 0.06$ mT (corresponds to $\alpha_\mathrm{0}=1.2\times 10^{-4}$ at a frequency $f=6.8$ GHz), whereas for the Pt covered YIG sample  $\Delta H_\mathrm{Pt}= 0.16$ mT (corresponds to $\alpha_\mathrm{Pt}=3.3\times 10^{-4}$, respectively). Thus, we obtain a change of the Gilbert damping constant $\Delta \alpha = 2.1\times 10^{-4}$. 
The spin mixing conductance \g\ is related to the change of the Gilbert damping $\Delta \alpha= \alpha_\mathrm{Pt}- \alpha_\mathrm{0}$ as\cite{Nakayama-2012,Mosendz,Heinrich}

\begin{equation}
\label{mixing}
g_\mathrm{eff}^{\uparrow \downarrow}= \frac{4\pi M_\mathrm{S} d_\mathrm{F}}{g \mu_\mathrm{B}} \Delta \alpha .
\end{equation}

Taking $M_\mathrm{S}= 140$~kA/m and $d_\mathrm{F}= 2.1$ \textmu m  into account we obtain a spin mixing conductance of \g$= 2.35 \times 10^{19}$ m$^{-2}$ for this particular sample. Since the spin mixing conductance \g\ is proportional to the spin pumping efficiency and, thus, consequently to the enhancement parameter $\bar{\epsilon}$, we can calculate \g\ for the other surface treatments. The results are summarized in Tab.~\ref{tab}.

As it is apparent from Tab.~\ref{tab}, the spin mixing conductance can be varied by treating the YIG surface before the Pt deposition in the range of two orders of magnitude. The largest value of the spin mixing conductance is obtained for a combined surface treatment by ``piranha''  etch and O$^{+}$/Ar$^{+}$ plasma, \g= 3.43$\times 10^{19}$~m$^{-2}$. The obtained values for \g\ agree with the values reported in the literature. \cite{Goennenwein,Rezende,Burrowes,Heinrich} Our maximal spin mixing conductance \g= 3.43$\times 10^{19}$~m$^{-2}$ is one order of magnitude larger than the one reported in Refs.~\cite{Burrowes,Heinrich} for YIG/Au (\g =5$\times 10^{18}$ m$^{-2}$) and even three orders of magnitude larger than the one estimated in Ref.~\cite{Kajiwara} for YIG/Pt (\g =3$\times 10^{16}$~m$^{-2}$). On the other hand, our maximal value is still one order of magnitude smaller compared to the one reported in  Ref.~\cite{Rezende} for YIG/Pt (\g =4.8$\times 10^{20}$~m$^{-2}$).

In conclusion, we have shown a strong dependence of the spin pumping effect on the interface condition of YIG/Pt bilayer structures. 
We improved the ISHE signal strength by a factor of more than 150 using a combination of ``piranha'' etch and \textit{in-situ} O$^{+}$/Ar$^{+}$ plasma treatment in comparison to standard ultrasonic cleaning. The combined cleaning by ``piranha'' etch and heating at $500^\circ$C yields a comparable enhancement of the spin pumping efficiency (by a factor of 104). The spin mixing conductances for the different surface treatments were calculated. We find a maximal value of \g= 3.43$\times 10^{19}$ m$^{-2}$. Since the voltage generated by the ISHE scales with the length of the Pt electrode, optimal interface conditions are extremely essential for the utilization of spin pumping and ISHE in micro-scaled devices. Our results are also important for studies on the reversed effects:  the amplification \cite{Wu2011} and excitation \cite{Kajiwara} of spin waves in YIG/Pt structures by a combination of the direct spin Hall and the spin-transfer torque effect. \cite{Kajiwara,Berger}

We thank E.~Saitoh and K.Ando for helpful discussions. Financial support by Deutsche Forschungsgemeinschaft (CH 1037/1-1) and the Nano-Structuring Center, TU Kaiserslautern, for technical support, is gratefully acknowledged.


\begin{thebibliography}{19}
\bibitem{modes}In the $U_\mathrm{ISHE}$ signal as well as in the microwave absorption spectra several modes are visible. They are slightly different for each sample and are identified as higher width modes and perpendicular standing thickness spin-wave modes. Since only the maximal voltage is used to determine the spin pumping efficiency, the mode structures of the the different samples are of minor interest for the present study.
\end{thebibliography}
\end{document}